\def \pom {{\hspace{ -0.1em}I\hspace{-0.2em}P}}
\def \GeV {{\rm GeV}}

\def \mb {{\rm mb}}

\documentstyle[11pt]{aipproc}

\begin{document}

\rightline{VAND-TH-97-06}

\title{Factorization and Non-factorization in Diffractive Hard Scattering}

\author{Arjun Berera}
\address{Department of Physics and Astronomy\\
Vanderbilt University\\
Nashville, Tennessee 37235 U.S.A.}

\maketitle

\begin{abstract}
Factorization, in the sense defined for inclusive hard scattering, is
discussed for diffractive hard scattering.  A factorization theorem
similar to its inclusive counterpart is presented for diffractive DIS.
For hadron-hadron diffractive hard scattering, in contrast to its
inclusive counterpart, the expected breakdown of factorization is
discussed.  Cross section estimates are given from a simple field theory
model for non-factorizing double-pomeron-exchange (DPE) dijet production
with and without account for Sudakov suppression.
\end{abstract}

\medskip

To appear in DIS 97 proceedings, ed. J. O. Repond

\medskip


In the general
phenomena of diffractive hard scattering, the initial proton in DIS
or both protons at hadron collider participate in a hard process involving
a very large momentum transfer, but one or at hadron colliders one or both
hadrons 
is diffractively scattered, emerging
with a small transverse momentum and the loss of a rather
small fraction of longitudinal momentum. 
In this talk I will discuss the factorization theorem of diffractive DIS
and its expected breakdown in hadron-hadron initiated diffractive hard
processes.

Following \cite{bersop2}, I will formulate the hypothesis of diffractive
factorization in two step.
In the first stage we hypothesize that the
diffractive structure function $F_2^{\rm diff}$ can be written in
terms of a diffractive parton distribution :
\begin{equation}
\frac{d F_2^{\rm diff}(\beta x_{\pom},Q^2;x_\pom,t)}{dx_\pom\,dt}
=x_{\pom} \sum_a \int_{\beta}^{1} d\beta '\
\frac{d\, f^{\rm diff}_{a/A}(\beta ' x_{\pom},\mu;x_\pom,t)}{dx_\pom\,dt}\
\hat F_{2,a}(\beta /\beta ',Q^2;\mu),
\label{factor}
\end{equation}
where $\hat F_2$ is the same function which is convoluted with the
inclusive parton densities to compute $F_2$ of inclusive DIS.
If for simplicity, we ignore $Z$ exchange, then 
%
%
$\hat F_{2,a}(\beta/\beta ',Q^2;\mu)
= e_a^2\ \delta(1-\beta/\beta ') + {\cal O}(\alpha_s)$.
%
%

In the second stage, we hypothesize that the diffractive parton distribution
function has a particular form:
\begin{equation}
\frac{d\,f^{\rm diff}_{a/A}(\beta x_{\pom},\mu;x_\pom,t)}{dx_\pom\,dt} 
={1 \over 8\pi^2}\, |\beta_{A}(t)|^2 x_\pom^{-2\alpha(t)}\,
f_{a/\pom}(\beta, t,\mu)\,.
\label{pomdist}
\end{equation}
Here $\beta_A(t)$ is the pomeron coupling to hadron A and $\alpha(t)$
is the pomeron trajectory.
The function $f_{a/\pom}(\beta, t,\mu)$ defined above is the
``distribution of partons in the pomeron''.
I distinguish the ``diffractive factorization'' of
Eq.~(\ref{factor}) from the ``Regge factorization''
of Eq.~(\ref{pomdist}).
The latter is a special case of the former.
The Ingelman-Schlein model \cite{IS} is synonymous with "Regge
factorization''.  The structure function 
$F_2^{diff}(\beta x_{\pom}, Q^2; x_{\pom}, t)$ for the IS-model
is obtained by inserting 
Eq.~(\ref{pomdist}) into
(\ref{factor}).  An inconsistency of data to the IS-model does not
also imply an inconsistency to diffractive factorization.

I now give operator definitions of the diffractive
parton distribution. The diffractive distribution of a quark of
type $j \in \{u,\bar u,d,\bar d,\dots\}$ in a hadron of type $A$ 
in terms of field operators $\tilde\psi(y^+,y^-,{\bf y})$
evaluated at $y^+ = 0$, ${\bf y} =0$ is:
\begin{eqnarray}
&&\frac{d\, f^{\rm diff}_{a/A}(\beta x_{\pom},\mu; x_{\pom}, t)}{
dx_\pom\,dt}  = 
{1 \over 64{\pi}^3}{1 \over 2}\sum_{s_{\!A}}\int d y^-
e^{-i\beta x_{\pom} P_{\!A}^+ y^-}
\nonumber\\
&& \sum_{X,s_{\!A^\prime}}
\langle P_{\!A},s_{\!A} |\tilde {\overline\psi}_j(0,y^-,{\bf 0})
| P_{\!A^\prime},s_{\!A^\prime}; X \rangle
\gamma^+ \langle P_{\!A^\prime},s_{\!A^\prime}; X|
{\tilde {\psi}}_j(0)| P_{\!A},s_{\!A} \rangle.
\label{fdiff1}
\end{eqnarray}
We sum over the spin $s_{\!A^\prime}$ of the final state proton and over
the states $X$ of any other particles that may accompany it.
Similarly, the diffractive distribution of gluons in a hadron is
\begin{eqnarray}
&&{d\, f^{\rm diff}_{a/A}(\beta x_{\pom},\mu; x_{\pom}, t)\over
dx_\pom\,dt} =
{1 \over 32{\pi}^3 \beta x_{\pom} P_{\!A}^+}{1 \over 2}\sum_{s_{\!A}} \int d y^-
e^{-i\beta x_{\pom} P_{\!A}^+ y^-}
\nonumber\\
&&\sum_{X,s_{\!A^\prime}}
\langle P_{\!A},s_{\!A} |\tilde F_a(0,y^-,{\bf 0})^{+\nu}
| P_{\!A^\prime},s_{\!A^\prime}; X \rangle
\langle P_{\!A^\prime},s_{\!A^\prime}; X|
\tilde F_a(0)_\nu^{\ +}| P_{\!A},s_{\!A} \rangle.
\label{fdiff2}
\end{eqnarray}
The proton state $| P_{\!A},s_{\!A} \rangle$ has spin $s_{\!A}$ and
momentum $P_{\!A}^\mu =  (P_{\!A}^+, {M_{\!A}^2 / [2P_{\!A}^+]},{\bf
0})$. We average over the spin. Our states are normalized to
%
%
$\langle k |p \rangle = 
(2\pi)^3\, 2p^+\,\delta(p^+ - k^+)\,\delta^2({\bf p} - {\bf k})$.
%
%
The tilde on the fields $\tilde \psi_j(0,y^-,{\bf 0})$
and $\tilde F_a(0,y^-,{\bf 0})^{+\nu}$ is to imply that they are
multiplied by an exponential of a line
integral of the vector potential as shown in \cite{bersop2}.

The diffractive parton
distributions are ultraviolet divergent and require
renormalization. It is convenient to perform the renormalization
using the $\overline{\rm MS}$ prescription, as
discussed in \cite{CSdistfns,CFP}. This introduces a renormalization
scale $\mu$ into the functions. In applications, one sets $\mu$ to be
the same order of magnitude as the hard scale of the physical process.

The renormalization involves ultraviolet divergent subgraphs.
Subgraphs with more than two
external parton legs carrying physical polarization
do not have an overall divergence.
Thus the divergent subgraphs are the same as for the ordinary parton
distributions. We conclude that the renormalization group equation
for the diffractive parton distributions is
\begin{equation}
\mu { d  \over d\mu}\,
{ d f^{\rm diff}_{a/A}(\beta x_{\pom},\mu;x_\pom,t) \over dx_\pom\,dt}=
\sum_b \int_{\beta x_{\pom}}^1 { d{z} \over {z}}\
P_{a/b}(\beta x_{\pom}/ z,\alpha_s(\mu))\ 
{ d f^{\rm diff}_{b/A}(z,\mu;x_\pom,t) \over dx_\pom\,dt}
\label{APeqn}
\end{equation}
with the same DGLAP kernel \cite{DGLAP},
$P_{a/b}( \beta x_{\pom} /z,\alpha_s(\mu))$, as one uses for the evolution of
ordinary parton distribution functions. 

The diffractive parton distribution ${d\,f^{\rm diff}_{a/A}
(\beta x_{\pom},\mu;x_\pom,t)/ dx_\pom\,dt}$, like the ordinary parton
distribution, is essentially not calculable using perturbative
methods. Recall, however, that it is possible to derive ``constituent
counting rules'' that give predictions for ordinary parton
distributions ${f_{a/A}(x,\mu)}$ in the limit $x \to 1$ for not too
large values of the scale parameter $\mu$ in the sense
of the analysis by Brodsky and Farrar \cite{brofar}. In the
same spirit, in \cite{bersop2} we have considered the diffractive
parton distributions in the limit $\beta \to 1$.

We find that the diffractive gluon distribution behaves as
$ (1-\beta)^p$
%
%
%
%
for $\beta \to 1$ at moderate values of the scale $\mu$, say 2
GeV, with
$0 \leq p \leq 1$.
%
%
%
%
The choice $p\approx 0$ corresponds to an effectively massless final
state gluon, while $p\approx 1$ corresponds to an effective gluon
mass.
For the diffractive quark distribution we find they behave as
$(1-\beta)^2$.
%
%
%
%
However, suppose that we interpret the calculation
as saying that the diffractive distribution of gluons is
proportional to $(1-\beta)^0$ for $\beta$ near 1 when the scale $\mu$ is
not
too large. Then the evolution equation for the diffractive parton
distributions will give a quark distribution that behaves like
\begin{equation}
{df_{q/A}^{\rm diff}(\beta x_\pom,\mu;x_\pom,t) \over {dx_\pom\,dt}}
\propto (1-\beta)^1,
\end{equation}
when the scale $\mu$ is large enough that some gluon to quark
evolution has occurred, but not so large that effective power $p$ in
$(1-\beta)^p$ for the gluon distribution has evolved substantially
from $p = 0$. A signature of this phenomenon is that the diffractive
quark distribution will be growing as $\mu$ increases at large
$\beta$, rather than shrinking. Perhaps this is seen in the data
\cite{ZeusH1}.

I will now turn to diffractive hard scattering in hadron-hadron
collisions.  There is an especially important difference in these
processes to their counterparts in the inclusive case.  For the latter
the leading twist cross section can be expressed as a product of
parton distribution functions, one for each hadron, 
times the hard partonic cross section.
Furthermore the parton distribution functions for the hadrons are the
same as those for inclusive DIS.  This is what we have understood as
factorization in inclusive hard processes.  In the diffractive case
factorization is expected to breakdown \cite{cfs,bersop,bercol}
when both impinging particles can interact strongly.  It is not expected
that the diffractive parton distributions of diffractive DIS should
correctly predict diffractive hard cross sections at hadron collider
nor does that appear to be true \cite{actw}.  Understanding the origin
of non-factorization challenges our theoretical knowledge of strong
interactions and in a fruitful direction since the effect is
experimentally measurable.

We have been studying non-factorization in a particular model for double
pomeron exchange (DPE) dijet production \cite{bercol}.  For the
remainder of this talk, I will report on that work.
The reaction of interest is
\begin{equation}
A+B\rightarrow A'+B'+\mbox{\rm 2 jets},
\label{DPE.jets}
\end{equation}
where hadrons $A$ and $B$ lose tiny fractions $x_a$ and $x_b$ of their
respective longitudinal momenta,
and they acquire transverse momenta ${\bf Q}_1$
and ${\bf Q}_2$.
Such events are called hard
double-pomeron exchange (DPE) events because both
incoming hadrons survive unscathed with a hard process in the central
region of final-state rapidity.
The process (\ref{DPE.jets}) has a quite 
dramatic signature: the final state consists
of the two diffracted hadrons, two high-$E_T$ jets, and {\em
nothing} else.

I present results from our work \cite{bercol}, that applied
the CFS mechanism \cite{cfs} in a simple field theory model to compute
the DPE jet cross section
with the lowest order Feynman
graphs that are appropriate.  
Although numerical estimates from our model in its base form
are crude,
the results of our calculation establish
that the exclusive processes of DPE to jets is leading twist and
non-factorizing.
This is quite non-trivial, since
firstly, there are in fact several two-jet emission graphs of
which only certain survive and
secondly, some of our graphs are a
power law larger than the final answer.  The proof of the necessary
cancelation relies on Ward identities and power counting \cite{bercol}.
To show that general
principles do not imply some other cancelation, it is important to
have a complete, consistent and gauge-invariant model, which ours
provides.

One modification to our base model in \cite{bercol} is to treat Sudakov
suppression \cite{kmr,coll1}. This effect arises from soft and collinear
gluon emission before the parton-parton hard vertex.  In an inclusive
hard process, this effect is not present due to cancelation of
appropriate real and virtual emissions.  However the diffractive
constraint prohibits real emissions along the direction of the
diffractive final state hadron.  I have computed the DPE dijet cross
section with and without the Sudakov suppression factor and the results
are in table 1.  The kinematic limits I used were $E_T > 5 {\rm GeV}$,
$0< x_a,x_b <0.05$ and I have integrated over jet rapidities $y_-$,
$y_+$ consistent with these cuts.   

\begin{table}
\begin{tabular}{c|cc}
 $\sqrt{s}$                 
&  \multicolumn{2}{c}{$\sigma_{dijet}^{NDPE}(E_T^{min}=5.0)$}   \\
\cline{2-3}
                      & {\rm without Sudakov}  & {\rm with Sudakov} \\
 {\rm (GeV)}                                & {\rm (mb)}  & {\rm (mb)} \\
\hline
630   & $0.044$ & $0.024$ 
 \\
1800   &  $0.17$ & $0.086$ 
\\
14000   &  $0.65$ & $0.31$  
\\
\end{tabular}
\caption{Table 1: Non-factorizing Double Pomeron dijet cross section
with and without one loop Sudakov suppression with 
$E_T^{min}=5 {\rm GeV}$.}
\end{table}

One can compare the results in
table 1 to
(a) the
inclusive two-jet cross section
(i.e., without a diffractive requirement:
$A+B\to\mbox{\rm jet}+\mbox{\rm jet}+X)$, and
(b) the result of applying the
Ingelman-Schlein model to DPE \cite{IS,bercol}, which gives a result for the
process $A+B\to\mbox{\rm $A'+B'+$jet}+\mbox{\rm jet}+X$. This process we call
factorized double-pomeron-exchange (FDPE).

For this I find the total cross
sections integrated over $y_{+}$, $y_{-}$ with the same $x_a$, $x_b$
cuts and for $E_T>5.0~\GeV$
are, at $\sqrt {s}=1800~\GeV$,
$\sigma_{\rm incl}(1800,5)=2.4~\mb$, 
$\sigma_{\rm FDPE}(1800,5)=0.0022~\mb$, 
and at $\sqrt {s}=630~\GeV$,
$\sigma_{\rm incl}(630,5)=0.31~\mb$,
$\sigma_{\rm FDPE}(630,5)=0.000062~\mb$.

There is no experimental cross sections yet reported for the DPE dijet
process.
However the preliminary CDF/D0 results \cite{cdf,d0} suggest that our
cross section estimates are too high perhaps even by a couple of orders
of magnitude.  We have not as yet treated absorptive corrections in our
calculation.  It remains to be seen how much of a suppression they will
give. It also remains to be seen what will be the experimentalists
final results.  Thus a test of our model still awaits further
developments.


\begin{references}
\bibitem{bersop2} A.\ Berera and D.\ E.\ Soper,  Phys. Rev. D {\bf 53},
6162, (1996).

\bibitem{IS}  G.\ Ingelman and P.\ Schlein,
{\em Phys.\ Lett.} B {\bf 152}, 256 (1985).

\bibitem{CSdistfns} J.\ C.\ Collins and D.\ E.\ Soper,
{\em Nucl.\ Phys.}  B {\bf 194}, 445 (1982).

\bibitem{CFP}
G.\ Curci, W.\ Furmanski and R.\ Petronzio,
{\em Nucl.\ Phys.} B {\bf 175}, 27 (1980).

\bibitem{DGLAP} V.\ N.\ Gribov and L.\ N.\ Lipatov,
{\em Sov.\ J.\ Nucl.\ Phys.} {\bf 15}, 438 (1972);
Yu.\ L.\ Dokshitzer, {\em Sov. Phys. JEPT} {\bf 56}, 641 (1977);
G.\  Altarelli and G.\  Parisi, {\em Nucl.\ Phys.} B {\bf 26}, 298 (1978).

\bibitem{brofar} 
S.\ J.\ Brodsky and G.\ Farrar,
{\em Phys.\ Rev.} D {\bf 11}, 1309 (1975).

\bibitem{ZeusH1} ZEUS Collaboration 
(M. Derrick, {\it et al.}), {\em Z. Phys.} C {\bf 70}, 391 (1996);
H1 Collaboration, (T.\ Ahmed {\it et al.}),
{\em Phys.\ Lett.} B {\bf 348}, 681 (1995).
 


\bibitem{cfs} J.C. Collins, L. Frankfurt, and M. Strikman,
{\em Phys.\ Lett.} B {\bf 307}, 161 (1993).

\bibitem{actw} L. Alvero, J. C. Collins, J. Terron, and J. Whitmore,
hep-ph/9701374.
 
\bibitem{bersop} A. Berera and D. E. Soper, Phys. Rev. D {\bf 50}, 4328
(1994).

\bibitem{bercol} A. Berera and J. C. Collins, Nucl. Phys. B {\bf 474}, 183
(1996).

\bibitem{kmr} V. A. Khoze, A. D. Martin and M. G. Ryskin,
hep-ph/9701419.

\bibitem{coll1} J. C. Collins, in {\it Perturbative Quantum Chromodynamics},
A. H. Mueller (ed.), 573 (1989).

\bibitem{cdf}  P. L. Melese, (CDF Collaboration), these proceedings.

\bibitem{d0} A. Brandt (D0 Collaboration), these proceedings.
\end{references}
\end{document}